\documentclass[final]{raa06}           
\usepackage{graphicx,times}             
\usepackage{natbib}
\usepackage{amssymb,amsmath}
\bibpunct{(}{)}{;}{a}{}{,}
\usepackage[colorlinks=true, citecolor=blue]{hyperref}%

\usepackage{graphicx,kantlipsum,setspace}
\usepackage{caption}
\usepackage{graphics,epsf}
\usepackage{amsmath}                
\usepackage{amsfonts}               
\usepackage{amssymb}                
\usepackage{epsfig}                 
\usepackage{appendix}
\usepackage{graphicx}
\usepackage{float}
\usepackage{color}
\usepackage{multirow}
\usepackage{colortbl}
\usepackage[para,online,flushleft]{threeparttable}
\usepackage{xcolor}

\hypersetup{citecolor=blue, 
            linkcolor=red, 
            menucolor=blue, 
            urlcolor=blue}  

\newcommand{\cm}{{~\rm cm}}

\newcommand{\km}{{~\rm km}}
\newcommand{\s}{{~\rm s}}

\newcommand{\erg}{{~\rm erg}}
\newcommand{\yr}{{~\rm yr}}

\newcommand{\keV}{{~\rm keV}}

\begin{document}

   \title{The Puppis A supernova remnant: an early jet-driven neutron star kick followed by jittering jets
}

   \volnopage{Vol.0 (20xx) No.0, 000--000}      
   \setcounter{page}{1}          


   \author{Ealeal Bear, Dmitry Shishkin, Noam Soker
    }

   \institute{Department of Physics, Technion, Haifa, 3200003, Israel;   {\it    ealeal44@technion.ac.il;  s.dmitry@campus.technion.ac.il; soker@physics.technion.ac.il}\\
\vs\no
   {\small Received~~20xx month day; accepted~~20xx~~month day}}

\abstract{
We identify a point-symmetric morphology of three pairs of ears/clumps in the core-collapse supernova remnant (CCSNR) Puppis A, supporting the jittering jets explosion mechanism (JJEM). In the JJEM, the three pairs of jets that shaped the three pairs of ears/clumps in Puppis A are part of a large, about 10 to 30 pairs of jets that exploded Puppis A. Some similarities in morphological features between CCSNR Puppis A and three multipolar planetary nebulae considered to have been shaped by jets solidify the claim for shaping by jets. 
Puppis A has a prominent dipole structure, where one side is bright with a well-defined boundary, while the other is faint and defused. The neutron star (NS) has a natal kick velocity in the opposite direction to the denser part of the dipole structure. We propose a new mechanism in the frame of the JJEM that imparts a natal kick to the NS, the kick-by-early asymmetrical pair (kick-BEAP) mechanism. At the early phase of the explosion process, the NS launches a pair of jets where one jet is much more energetic than the counter jet. The more energetic jet compresses a dense side to the CCSNR, and, by momentum conservation, the NS recoils in the opposite direction. Our study supports the JJEM as the primary explosion mechanism of core-collapse supernovae and enriches this explosion mechanism by introducing the novel kick-BEAP mechanism.
}

\maketitle

\section{Introduction}
\label{sec:Introduction}

In the jittering jets explosion mechanism (JJEM; \citealt{PapishSoker2011}) of core-collapse supernovae (CCSNe), pairs of opposite jets with stochastically (fully or partially) varying directions explode the star. 
The newly born neutron star (NS)—or the black hole, in cases where the NS collapses into one—launches several to a few tens of pairs of jets on a timescale of $\approx 0.5-10 \s$, as it accretes mass from the collapsing core via intermittent accretion disks (e.g., \citealt{Soker2024Keyhole}); later pairs of jets are possible. In the present study, we will propose that the first jets are launched already at $t_{\rm b} \simeq 0.1-0.2 \s$ post-bounce (Section \ref{sec:kick}). According to the JJEM, the stochastic angular momentum of the gas that feeds the intermittent accretion disks results from instabilities above the NS that increase the angular momentum seed fluctuations of the collapsing gas (e.g., \citealt{ShishkinSoker2021}).
These instabilities, occurring above the NS and below the stalled shock at $\simeq 150 \km$ from the center, include modes of the spiral standing accretion shock instability (e.g.,  \citealt{Buelletetal2023} for a recent study of this instability). The seed fluctuations come from the convective motion in the pre-collapse core (e.g., \citealt{PapishSoker2014Planar, GilkisSoker2014, GilkisSoker2016, ShishkinSoker2023, WangShishkinSoker2024}). 
Neutrino heating enhances the explosion process in the JJEM; however, the jets remain the primary drivers of the explosion \citep{Soker2022nu}. Jets dominate the process even when an energetic magnetar contributes additional post-explosion energy (e.g., \citealt{SokerGilkis2017, Kumar2025}).

In the competing delayed neutrino explosion mechanism of CCSNe (e.g., \citealt{Andresenetal2024, Burrowsetal2024kick, JankaKresse2024, Muler2024, Mulleretal2024, Nakamuraetal2024, vanBaaletal2024, WangBurrows2024, Laplaceetal2024, Huangetal2024}, for some very recent studies) neutrino heating revives the stalled shock and explodes the star; jets play no roles in the explosion process. The jet-based magnetorotational explosion mechanism requires a rapidly rotating pre-collapse core to launch jets along a fixed axis (e.g., \citealt{Shibagakietal2024, ZhaMullerPowell2024}, for recent studies of this mechanism); therefore, it can power only a very small fraction of CCSNe. We consider the magnetorotational explosion mechanism part of the neutrino-driven mechanism despite the powering by jets because it still attributes most CCSNe to the neutrino-driven mechanism. The JJEM attributes all CCSN explosions to jets. 

The observational property of CCSN remnants (CCSNRs) that best differentiates between the two above-mentioned explosion mechanisms is point-symmetric morphological features. A point-symmetric morphology has two or more pairs of structural features on opposite sides of the center that do not share the same symmetry axis. The opposite structures include clumps, filaments, bubbles (faint, enclosed structures with brighter rims), lobes (bubbles with partial rims), and ears. 

While the JJEM predicts that many, but not all, CCSNR morphologies are point symmetric, the neutrino-driven explosion mechanism has no explanation for most of these morphologies (for a detailed comparison between the two explosion models, see \citealt{SokerShishkin2024Vela}). The following is the list of CCSNRs with studies that attributed their point-symmetric morphologies to the JJEM (see also \citealt{Soker2025Two}): 
SNR 0540-69.3 \citep{Soker2022SNR0540},
Vela (\citealt{Soker2023SNRclass, SokerShishkin2024Vela}), 
CTB~1 \citep{BearSoker2023RNAAS}, 
N63A \citep{Soker2024CounterJet}, 
SN 1987A \citep{Soker2024NA1987A, Soker2024Keyhole}, 
G321.3–3.9 \citep{Soker2024CF, ShishkinSoker2024}, 
G107.7-5.1 \citep{Soker2024CF}, 
Cassiopeia A \citep{BearSoker2024}, 
the Cygnus Loop \citep{ShishkinKayeSoker2024}, 
W44 \citep{Soker2025W44}, 
and the Crab Nebula \citep{ShishkinSoker2025Crab}.
Some other CCSNRs show only one pair of opposite ears that were analyzed in the frame of the JJEM \citep{GrichenerSoker2017}, e.g., SNRs G309.2-00.6 and 3C58 (see images of these sources in \citealt{Gaensleretal1998} and \citealt{Slaneetal2004}).

Many of the exploding jets leave no observable marks on the CCSNR. This study considers the jets that do leave marks on the CCSNR. The shaping jets form, among others, pairs of ears, a main line of symmetry (main jet axis), and point-symmetric morphologies in many CCSNRs. The similarities of these, and others, morphological structures to similar structures in planetary nebulae (e.g., \citealt{Bearetal2017, BearSoker2017, BearSoker2018, Soker2022Rev, Soker2024PNSN}) and cooling flow clusters \citep{Soker2024CF} support the claim of shaping by energetic jets. 
Several properties, like the large volume of some jet-shaped structures that indicate energetic jets and the high abundance of heavy elements in some jet-shaped structures, show that the NS launches the shaping jets during the explosion process \citep{SokerShishkin2024Vela}. 

In general, studies of CCSNRs can teach us a lot about the explosion and the interaction with the interstellar medium (ISM) or circumstellar material (CSM; e.g., \citealt{SunGaoetal2022, Leietal2024, LuoTangMo2024, ShenBaoZhang2024}). We aim to study one CCSNR. 
This study analyzes the highly asymmetrical CCSNR Puppis A, which exhibits a complicated structure lacking clear symmetry. Earlier studies have argued for the jet-shaping of Puppis A. 
 \citet{Castellettietal2006} noted that the morphology of Puppis A bears a striking resemblance to that of the SNR W50, whose shell is distorted by precessing jets emanating from the central compact source SS433.  This resemblance suggests that jet activity may play a role in shaping Puppis A's morphology. \citet{GrichenerSoker2017} considered shaping by a pair of jets that are part of the jittering jets that exploded the progenitor of Puppis A and estimate the energy of each of the two jets that inflated the western and eastern ears to be $\simeq 1 \%$ of the explosion energy. Clearly, these two jets could not have by themselves exploded the star. We expect more pairs of jets, as we discuss in this study. Other studies noticed the interaction of Puppis A with a CSM or ISM (e.g., \citealt{DubnerGiacani2015, ReynosoWalsh2015, Meyeretal2022}).  However,  \citet{Reynosoetal2017} find no strong indication for ejecta-cloud interaction in the east (they do find possible indication for interaction in the northern side). In any case, although the ISM influences the morphology of old CCSNRs (e.g., \citealt{Sofue2024}), ages $\gtrsim 1000 \yr$, it cannot account for point-symmetry and asymmetrical metal distribution as observed in many CCSNRs (e.g., \citealt{SokerShishkin2024Vela}). 

In this paper, we aim to further investigate the role of jets in shaping Puppis A through a detailed analysis of its X-ray and infrared emission. We describe its structure as appears in observations of earlier studies but emphasize its point-symmetrical morphology and highly asymmetrical dipole structure in Section \ref{sec:MorphologicalFeatures}. In Section \ref{sec:PlanetaryNebulae}, we find some similarities with planetary nebulae that strengthen the claim for shaping by jets. In Section \ref{sec:kick}, we propose a novel mechanism for the kick velocity unique to the JJEM, the kick by early asymmetrical pairs (Kick-BEAP) mechanism. We summarize this study in Section \ref{sec:Summary}. 
 
\section{The point-symmetric structure of Puppis A}
\label{sec:MorphologicalFeatures}

Puppis A is a CCSNR projected inside the Vela CCSNR. Its morphology and structure are highly non-spherical, and it has attracted attention over the years (e.g., \citealt{Arendtetal1990,  Arendtetal2010, Hwangetal2005, Castellettietal2006, Katsudaetal2010, Hewittetal2012, Dubneretal2013, Reynosoetal2017, Reynosoetal2018, Arugaetal2022, Krivonosetal2022, Mayeretal2022, Giuffridaetal2023, Ghavamianetal2024}). In this study, we concentrate on two aspects of its morphology. (1) We reveal a possible point-symmetric structure of three pairs: two pairs of `ears' and one pair of clumps. An `ear' is a protrusion from the SNR's central nebula with a base smaller than the main SNR and a cross-section that decreases outward. The ear might appear differently from the main SNR, like being fainter. Our method explicitly requires the occurrence of two nearly diametrically opposed features.
(2) Motivated by the prominent dipole asymmetry, we propose a new mechanism to impart a kick velocity to the newly born NS, as further investigated in Section \ref{sec:kick}. 

\begin{figure*}
\begin{center}
\includegraphics[trim=2cm 1cm 3cm 2cm, clip, width=\textwidth]
{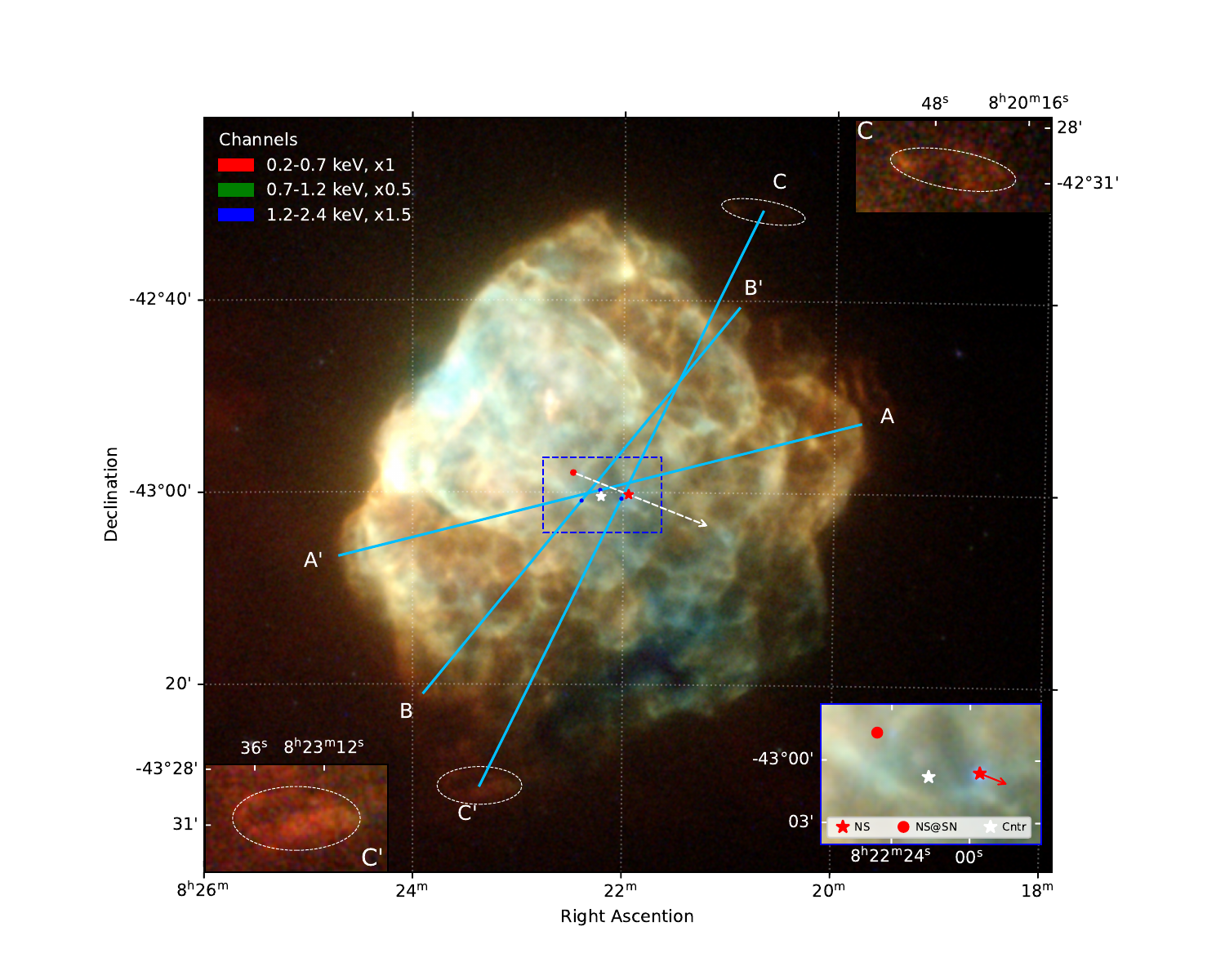}
\caption{X-ray log scaled image of Puppis A in the $0.2-2.4 \keV$ range taken from the eROSITA EDR data  (for an early eROSITA image see \citealt{Mayeretal2022}).  Colors indicate different energy bands, with levels balanced for display purposes (see the legend in the upper left). Light-blue lines indicate two pairs of point-symmetric ears (AA$^\prime$, BB$^\prime$) and one pair of clumps (CC$^\prime$).  The two dashed-white ellipses indicate the location of clumps C and C$^\prime$, but not necessarily their shape.   We denote the center of each symmetry line (axis) with a blue dot and the total average projected location of the three centers with a white asterisk. We mark the present NS location with a red asterisk. 
A dashed white arrow shows the direction of the NS’s proper motion, while a red dot marks the estimated location of the NS at the time of the explosion \citep{Mayeretal2020}. 
In the inset on the bottom right, we display a zoom-in on the central region of Puppis A; the red arrow indicates the NS's proper motion direction.
In the upper-right and lower-left insets, we zoom on clumps C and C$^\prime$, respectively.  }
\label{fig:PuppisAFig1}
\end{center}
\end{figure*}

In our analysis, we make use of X-ray observations of Puppis A, using data from the early data release (EDR) of eROSITA (\citealt{Predehl_eROSITA_2021}; ObsID: 700195). Data were processed using the eROSITA science analysis software system (eSASS; \citealt{Brunner_eSASS_2022}). Using the \texttt{evtool} software with default settings, we extracted individual energy bands to construct the X-ray image, which is shown in Figure \ref{fig:PuppisAFig1}.
We added three lines connecting two pairs of ears: A with A$^\prime$, B with B$^\prime$, and one pair of clumps C with C$^\prime$.Earlier studies identified the pair AA$^\prime$ (e.g., \citealt{Castellettietal2006}). \citet{GrichenerSoker2017} connect each ear to the NS rather than the opposite ear. This forms a bend in the direction of the two ears. Namely, as seen from the NS, the two ears are not exactly opposite at  $180^\circ$ to each other. We prefer to connect the pair AA$^\prime$ with a straight line because we consider the NS kick velocity that takes the NS away from the explosion location. Inhomogeneities in the CSM or the ISM with which the outer ejecta interacts, if they exist, might change the ear morphology in a tangential direction, hence, displacing its center tangentially. Therefore, there are some uncertainties in the exact directions of the original jets. Connecting by straight lines is the most straightforward approach to identifying a point-symmetric structure.   
We also identify the ear pair BB$^\prime$ and clump pair CC$^\prime$. 
The northern parts of these two pairs overlap on the plane of the sky but are not parts of a single structure. 
We take it to imply that the two axes (lines) of pairs BB$^\prime$ and CC$^\prime$ are highly inclined to each other.    

In identifying the two pairs of ears AA$^\prime$ and BB$^\prime$ we apply the following rules. ($i$) The protrusion has a cross-section that decreases outward. ($ii$) Most of the protrusion area has a different surface brightness than the main part of the SNR in at least one of the emission bands. The bright zone above ear A$^\prime$ in the east (called ``the bright eastern knot'' or ``BEK'', by \citealt{Petreetal1982}) is not a protrusion with a detached structure from the main CCSNR; hence, we do not classify it as an ear. ($iii$) We agree that the protrusion can be classified as an ear based on the ears of planetary nebulae (see Section \ref{sec:PlanetaryNebulae}).  
We note other protrusions in the images of Puppis A but do not find them as robust ears (see below for the two potential protrusions in the west). Future deeper X-ray observations with high spatial resolution might reveal more pairs.  

 In pairing ears, we note that each ear has a more or less symmetry axis; it goes from the tip of the ear to the center of its base on the main shell of the CCSNR. The symmetry axes point more or less to the center of the CCSNR. Therefore, the correct way is to connect opposite ears. The ears in a pair and the axes are not always exactly 180 degrees from each other because of the interaction of the ejecta with a CSM and/or an ISM, the NS kick, and instabilities in the explosion process \citep{SokerShishkin2024Vela}.  

We mark three central points on the image of Puppis A in Figure \ref{fig:PuppisAFig1}: the present location of the NS (red asterisk), the average location of the three centers of the three symmetry lines (white asterisk), and the estimated location of the NS at the time of the explosion (red dot). We determine the location of the NS at explosion from the proper velocity of the NS of $v_{\rm kick} = 763 \pm 73 \km \s^{-1}$ (\citealt{Mayeretal2020}; \citealt{HollandAshfordetal2017} give $v_{\rm kick} \simeq 437 \km \s^{-1}$), and an assumed SNR age of $4600 \yr$ (\citealt{Mayeretal2020}; \citealt{Arugaetal2022} give an age of $\simeq 10^4 \yr$). 
 As these numbers indicate, there are discrepancies in the velocity of the NS and the CCSNR age. Here, we use the more recently determined kick velocity and take the age from the same paper. We get a similar NS displacement to the above value if we take the kick velocity from \citet{HollandAshfordetal2017} and the age from \citet{Arugaetal2022}.  The differences between different estimated NS original locations are within the other uncertainties of the point-symmetric structure; hence, they do not matter much to our study.   

Although the three symmetry lines do not cross at the same point, they all cross between the initial (at the explosion time) and the present location of the NS. Puppis A's prominent dipole structure reinforces the robustness of the point-symmetric morphology.

\begin{figure*}
\begin{center}
\includegraphics[trim=0.95cm 2cm 0.95cm 3.4cm, clip, width=\textwidth]
{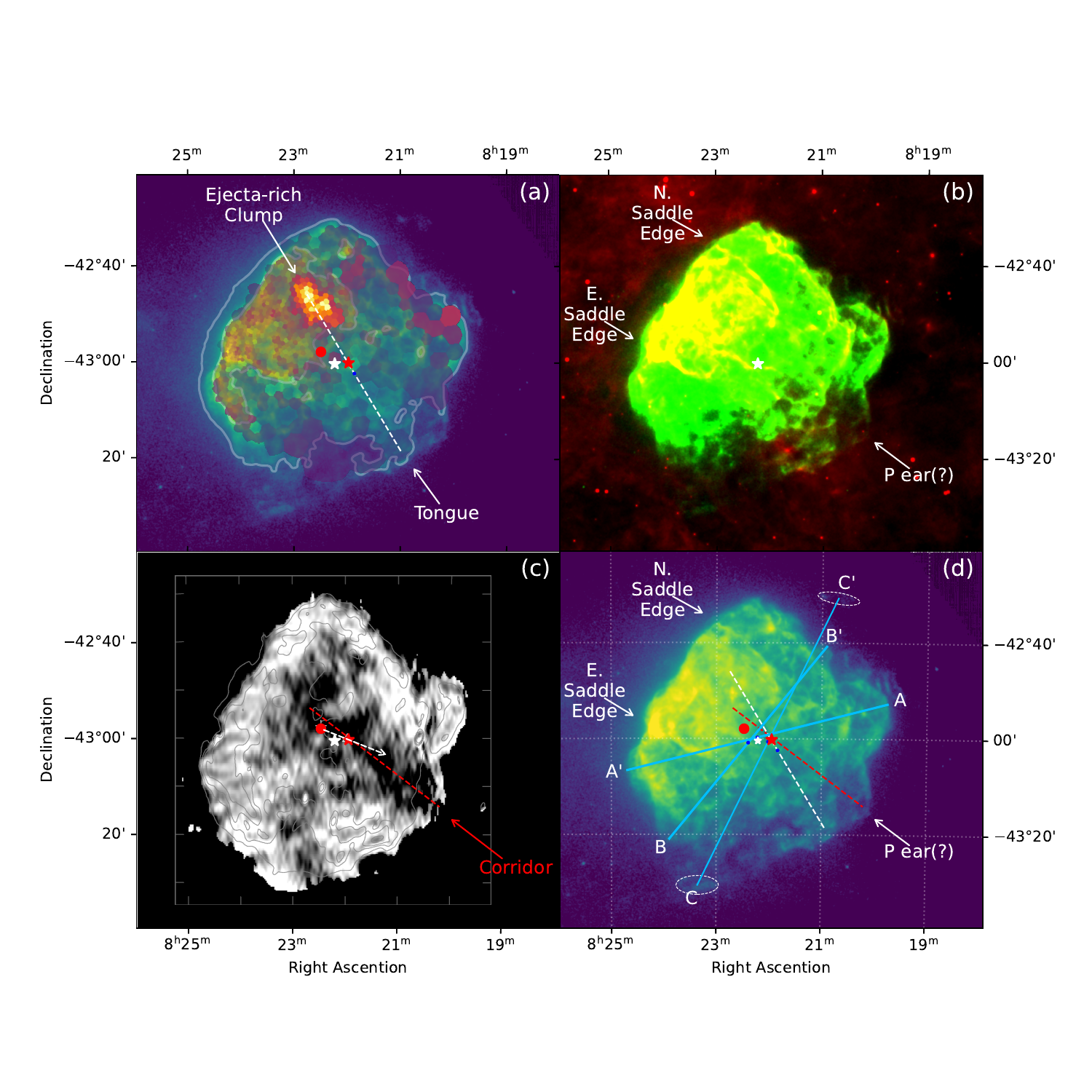}
\caption{Puppis A morphological features. (a) X-ray image (viridis colormap), with the Si/H abundances map in the range of 0 (transparent, originally: purple) to 3 (yellow) adapted from \citet{Mayeretal2022}. We denote the ejecta-rich clump, also enriched in O, Ne, Mg, S, and Fe (see \citealt {Mayeretal2022} for details). We mark a southwest protrusion that we call the ``tongue''. We draw a dashed white line between the ejecta-rich clump and the tongue, suggesting their potential formation in a pair of jets deserves further study. Two contours tracing X-ray emission lines outline the entire remnant and mark the brighter X-ray emitting regions in the northeastern part of Puppis A. (b) X-ray counts (green) and infrared (IR) image (WISE, $22\mu$m, red, from the All-WISE data release procured with \textit{SkyView}). The yellow region at the northeast part of the Puppis A SNR is where emission is strong in both X-ray and IR - which we nickname the ``Saddle''. We mark its northern and eastern edges. We also point at a potential ear (P-ear) that might be a counterpart to the saddle on the southwestern part of Puppis A. (c) Distribution of the radio continuum spectral index calculated between 327 and 1425 MHz from \citet{Castellettietal2006}, in grayscale. A dashed red line marks the darker central band in the image, corresponding to a flatter radio spectrum (see \citealt{Castellettietal2006} for details); we denote it the ``Corridor''. We also mark the NS kick direction (dashed white arrow), highlighting the small angle between them. Contour lines delineate the 1425 MHz emission from the original image by \citet{Castellettietal2006}. (d) X-ray counts image with all aforementioned relevant features we identified in Puppis A and the point-symmetric axes (these axes form a structure termed point-symmetric wind-rose) from Figure \ref{fig:PuppisAFig1}. A grid serves to guide the eye with the different features. In all panels, we note our proposed explosion center (see Figure \ref{fig:PuppisAFig1}) with a white asterisk, and in panels a, c, and d, the NS location (red asterisk) and the calculated location of the NS at explosion (red dot).}
\label{Fig:4panels}
\end{center}
\end{figure*}

In Figure \ref{Fig:4panels}, we present Puppis A as seen in different wavelengths. The dipole structure is clearly evident in panel b, with the X-ray bright emission region in the northeast displaying a relatively straight and sharp boundary. The bright zone forms a distinct saddle-like shape (hereafter called the Saddle); it is seen as a yellow-colored zone on panel b of Figure \ref{Fig:4panels}. On the opposite side of the dipole structure, the southwest side of Puppis A does not have a sharp boundary and appears much fainter.
The southwest's farthest structure is a faint potential ear we denote as the `P-ear' (`P ear(?)' in panels b,d in Figure \ref{Fig:4panels}). The question mark signifies that we are unsure whether this is a jet-shaped ear. Determining whether the southwest-most distant structure is a potential ear falls outside the scope of the current analysis. 

Southeast of the P-ear, we notice a zone that is brighter than the P-ear in X-ray (panels a,b,d in Figure \ref{Fig:4panels}) as seen most clearly by the contour line in panel a in Figure \ref{Fig:4panels}; we term this structure the `tongue.' The bright saddle on one side of Puppis A and the faint P-ear and the tongue on the other are the prominent components of the dipole morphology of Puppis A. 

\citet{Castellettietal2006}   present maps of the distribution of the radio continuum spectral index calculated between 327 and 1425 MHz, ranging $-0.8$ to $-0.2$ (figures 5 and 6 in their paper). These images show a central band from northeast to southwest with spectral index values that are flatter than their surroundings ejecta, having a radio spectral index of $\simeq -0.2$; this is the `corridor', which we denote on panel c of Figure \ref{Fig:4panels}. \citet{Castellettietal2006} suggest that the flatter spectrum might result from a continuous injection of relativistic particles by the NS.  
The NS's motion from explosion to present (proper motion) is inside and almost along the corridor. 

The eROSITA observations by \citet{Mayeretal2022} reveal a zone rich in ejecta material, namely, elements from the deep core of the progenitor, O, Ne, Mg, Si, S, and Fe. \citet{Mayeretal2022} mark in this zone the ejecta knot that was noticed by  \citet{Katsudaetal2008} and the ejecta-rich region that was seen by \citet{Hwangetal2008}. We mark this zone on panel a of Figure \ref{Fig:4panels}. Some studies attribute ejecta-rich clumps to jets and expect a symmetric structural feature on the other side of the center, as expected in the JJEM. A prominent example is the Vela SNR \citep{SokerShishkin2024Vela}. There is no ejecta-rich zone opposite the one northeast of Puppis A. However, we note that the tongue is on the opposite side of the ejecta-rich clumps, as we mark by the dashed-white line on panels a and d of Figure \ref{Fig:4panels}. At this time, we do not argue that the ejecta-rich zone and the tongue are opposite structural features formed by a pair of opposite jets. We only comment on the need for deeper observations to reveal the composition of the tongue and further explore the structure of the ejecta-rich clump. 

Some studies attribute the shaping to the wind of the progenitor of Puppis A, e.g., \citet{Reynosoetal2018}. There is also a dense ISM around Puppis A (e.g., \citealt{Arugaetal2022}). We do not question whether pre-explosion blown CSM and the ISM play a role in shaping Puppis A. However, we also attribute substantial shaping to pairs of jets that exploded the progenitor of Puppis A. We base our claim on the two pairs of ears  and one pair of clumps  that form the point-symmetric morphological structure. 
In Section \ref{sec:PlanetaryNebulae}, we find morphological similarities between Puppis A and a few planetary nebulae that are thought to be shaped by jets. This further supports the claim for jet-shaping of Puppis A.    

\citet{Ghavamianetal2024} propose the existence of a close companion to the progenitor of Puppis A, which likely shaped a funnel-like structure in the ejecta. This hypothesis aims to account for the blue-shifted, nested series of optically emitting rings observed near the center of Puppis A, referred to as “the Swirl”. We note that a binary companion might smear a point-symmetric morphology, as many other processes do, like interaction with a CSM and the ISM, the hot ejecta, instabilities during the explosion process, and the NS natal kick. 
 The CSM and ISM mainly affect the outer regions of the ejecta and might even erase small protrusions from the main ejecta shell, like small ears and weak jets. For these smearing processes, some SNRs lack a discernible point-symmetric morphology, or their symmetry becomes exceedingly difficult to identify. Consequently, a straightforward statistical analysis of how many CCSNRs exhibit recognizable point-symmetric morphology is unlikely to provide meaningful insights into the explosion mechanism.    

Pre-explosion mass loss by the progenitors of CCSNe might form a CSM with point-symmetric morphology. Spherical ejecta interacting with such a point-symmetric CSM can also result in point-symmetric SNRs, as suggested for the Type Ia SNR G1.9+0.3 \citep{Soker2024G19}. However, for CCSNRs, this explanation fails to account for key properties, such as isotropic ejecta (e.g., magnesium and neon) exhibiting point symmetry, point symmetry in inner regions of SNRs that have not yet interacted with the CSM, and small clumps with point-symmetric morphology that are unlikely to form through ejecta-CSM interactions (see a detailed study by \cite{SokerShishkin2024Vela}).
In Puppis A, the three symmetry axes (panel d in Figure \ref{Fig:4panels}) are to the southwest of the explosion location (red dot). 
This displacement aligns with the direction of the kick velocity, which is expected if the jets were launched after the NS had acquired its kick. Point-symmetric shaping by the CSM cannot account for this displacement, as the CSM was expelled before the NS gained its kick velocity.

Given that planetary nebulae often exhibit similar point-symmetric jet-driven morphologies, we compare Puppis A's morphology to well-studied examples of planetary nebulae to further support our claims. 

\section{Hints from planetary nebulae}
\label{sec:PlanetaryNebulae}

The method of identifying point-symmetric morphologies and attributing them to jets is widely used by researchers studying planetary nebulae (e.g., \citealt{Sahaietal2011, Clairmontetal2022, Danehkar2022, MoragaBaezetal2023, Mirandaetal2024, Sahaietal2024}). In this work, we apply the same approach

\begin{figure*}[t]
\begin{center}
\includegraphics[trim=1cm 0cm 4.5cm 1cm,width=\textwidth]
{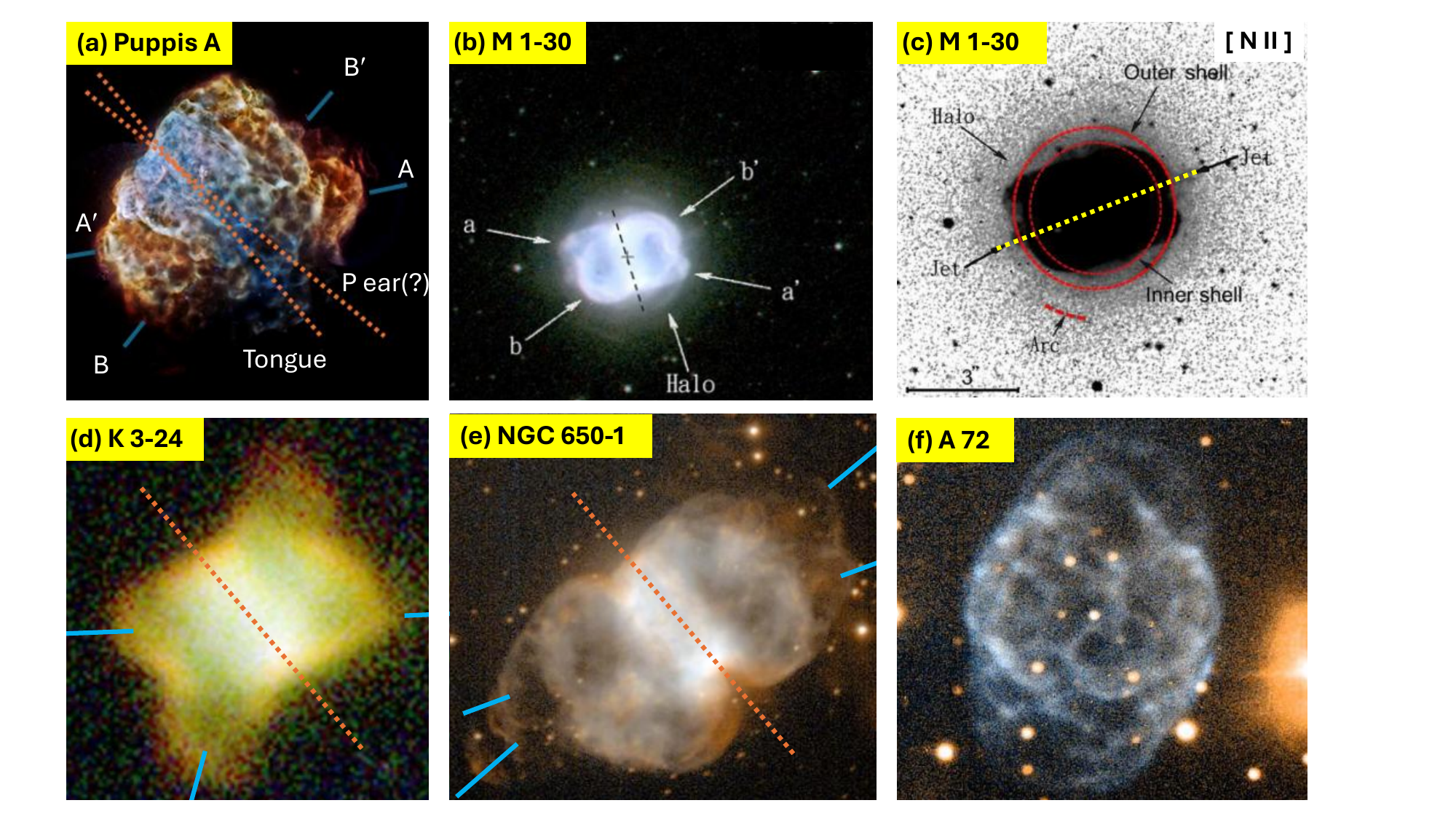}
\caption{Comparison of Puppis A’s morphology to planetary nebulae. Panel a, X-ray image of SNR Puppis A (Credit$^{\#}$: NASA/CXC/IAFE/, \citealt{Dubneretal2013}).  This X-ray image intensity scale emphasizes the granular texture of Puppis A.  We added the cyan lines to mark the two pairs of ears AA$^\prime$ and BB$^\prime$ (see Figure \ref{fig:PuppisAFig1}). Dotted orange lines mark a plane for comparison with the equatorial plane of the planetary nebulae; one line ends at the P-ear(?), and the other ends at the tongue (see Figure \ref{fig:PuppisAFig1}). 
Panels b and c: Two images of the multipolar planetary nebulae M~1-30 (PN G$355.9-04.2$) and marks from \citet{Hsiaetal2014}. We have added only the dotted yellow line connecting the two jets that they marked. The position of the central star is marked with a black cross. We consider the plane that \citet{Hsiaetal2014} marked by a black dotted line, the equatorial plane.
\citet{Hsiaetal2014} marked the two pairs of ears by aa' and bb'.
Panels d, e and f, are images of three different planetary nebulae from the IAC morphological catalog of northern Galactic planetary nebulae \citep{Manchadoetal1996}$^{\&}$: PN K~3-24 (PN~G$048.7+02.3$),  NGC~650-1 (PN G$130.9-10.5$), and PN A~72 (PN G$059.7-18.7$) respectively. In panels d and e, we added cyan lines to mark pairs of ears and dotted-orange lines to mark a possible equatorial plane of each planetary nebula.
\newline${\#}$ https://www.nasa.gov/image-article/unprecedented-x-ray-view-of-supernova-remains/
\newline \& Images taken from the PNIC catalog \citet{Balick2006}:  ${\rm https://faculty.washington.edu/balick/PNIC/}$
}
\label{fig:PuppisA4PNe}
\end{center}
\end{figure*}

In Figure \ref{fig:PuppisA4PNe}, we present Puppis A alongside four planetary nebulae with some, but not full, morphological similarities to Puppis A. The two images of M 1-30 are taken from \citet{Hsiaetal2014}.  This planetary nebula has an enhanced density in the equatorial plane with two lobes on its sides and a prominent pair of ears, aa', and a less prominent pair, bb'. We find a similarity between the pair aa' in M~1-30 and AA$^\prime$ in Puppis A. Panel c in Figure \ref{fig:PuppisA4PNe} is an overexposed image from  \citet{Hsiaetal2014} where they identify a pair of opposite jets (see their marks on the image). The axis of these two jets, the dashed-yellow line that we added on panel c, is perpendicular to the dense strip that  \citet{Hsiaetal2014} mark with a dashed-black line (Figure \ref{fig:PuppisA4PNe}b), which we take to be the equatorial plane. We find a similarity between this pair of jets in M~1-30 and the pair BB$^\prime$ in Puppis A, which is also perpendicular to a dense stripe, more or less the saddle-tongue structure. 
We mark two possible equatorial planes on the Puppis A image with dotted-orange lines (see Figure \ref{fig:PuppisA4PNe}a): one line extending to the P-ear(?) and the other reaching the tongue (see Figure \ref{Fig:4panels} for definitions).
A plane situated somewhere between the structures B and B$^\prime$ could be considered an equatorial plane solely for morphological comparison with planetary nebulae. However, in the case of the SNR Puppis A, it does not represent the equatorial plane of a binary system. The planetary nebula M 1-30 belongs to the class of multipolar planetary nebulae, which most researchers attribute to shaping by two or more pairs of jets. We propose a similar mechanism for Puppis A.

In the lower panels of Figure \ref{fig:PuppisA4PNe}, we present the planetary nebulae K~3-24 (panel d), NGC~650-1 (panel e), and A~72 (panel f) for its granular texture, originally from \citet{Manchadoetal1996}. The multipolar planetary nebulae K~3-24 and NGC~650-1 exhibit two pairs of ears, which we mark by pairs of light-blue lines and a dense equatorial plane between them (orange-dotted line). These features are also evident in SNR Puppis A, with a bright pair AA$^\prime$ and two pairs more or less perpendicular to the dense strip, i.e., pairs BB$^\prime$ and CC$^\prime$ (for pair CC$^\prime$ see Figure \ref{fig:PuppisAFig1}). 
The morphological similarities between Puppis A and the planetary nebulae K 3-24 and NGC 650-1 support that Puppis A may also be shaped by jets.

Notably, the planetary nebula A72, displayed in Figure \ref{fig:PuppisA4PNe}f, exhibits a granular morphology similar to the X-ray emission observed in Puppis A, composed of tens of cells (or grains). This morphological similarity suggests that the granular texture does not result from the explosion process or post-explosion radioactive heating (nickel bubbles). In planetary nebulae, the instability might result in the interaction of the main nebula with a previously ejected CSM or from a fast wind blown by the central star that accelerates the nebula (e.g., \cite{DeMarcoetal2022}). The same might hold in CCSNRs, where the ejecta interaction with a CSM might lead to granular texture, as the simulations by \citet{Orlandoetal2022} suggest for Cassiopeia A.  The granular structure appears in only a fraction of planetary nebulae and CCSNRs. Hydrodynamical simulations may further investigate the conditions that allow the development of a granular structure. We estimate that this requires the development of the relevant instabilities to a highly non-linear phase.

\section{Early jet-driven NS kick: The Kick-BEAP mechanism}
\label{sec:kick}

 The subject of this paper is the shaping of CCSNRs by jittering jets, focusing on CCSNRs with a pronounced dipole morphological component. The almost exact alignment between the NS kick velocity and the dipole direction of Puppis A (panel c of Figure \ref{Fig:4panels}) triggered the search for a relation between pairs of jets in the JJEM and the kick velocity. We use Puppis A's prominent point symmetry and dipole morphology to propose a new kick mechanism. Spin-kick alignment is observed in many pulsars (e.g., \citealt{Johnstonetal2005, Noutsosetal2012, BiryukovBeskin2025}).

In the past, JJEM studies considered the tug-boat mechanism to impart the natal kick to the NS. The tug-boat mechanism was developed in the frame of the delayed neutrino explosion mechanism (e.g., \citealt{Schecketal2004, Schecketal2006, Nordhausetal2010, Nordhausetal2012, Wongwathanaratetal2010, Wongwathanaratetal2013kick, Janka2017}); there are other natal kick mechanisms (e.g., \citealt{YamasakiFoglizzo2008, Yaoetal2021, Xuetal2022, LambiasePoddar2025}; for general discussion of natal kick velocities see, e.g., \citealt{Igoshev2020}). We here also take the mechanism of the kick velocity to be hydrodynamical, as evident, for example, from the anti-correlation between the direction of motion of the primary X-ray emitting ejecta and the neutron star observed in Cassiopeia A (e.g., \citealt{HwangLaming2012}).

Studies of 14 CCSNe show that the distribution of angles between the jet-main axis in the frame of the JJEM and the NS natal kick avoids small angles \citep{BearSoker2018kick, BearSoker2023RNAAS,  Soker2022SNR0540}. With the tug-boat mechanism, this implies that the angle of the main-jet axis to the dense clump that accelerates the NS avoids small angles. The first possible explanation holds that the jets prevent the formation of dense clumps along their propagation directions, and therefore, no NS acceleration takes place in those directions. The second explanation has several dense clumps that are falling towards the NS. Some clumps feed an accretion disk around the NS, and some escape and accelerate the NS by pulling it. This accretion disk launches jets along the angular momentum axis, perpendicular to the direction of the clumps' inflow, hence to the kick velocity. 

The gravitational tug-boat mechanism operates for a relatively long time of $\simeq 0.5-5 \s$ (e.g., \citealt{Wongwathanaratetal2013kick}). It converts the internal energy of the ejecta into kinetic energy of the ejecta, maintaining shell expansion \citep{Wongwathanaratetal2013kick}. Below, we suggest a very short mechanism to give the NS a kick velocity early in the explosion process. The tug-boat mechanism can operate later as well. 

The characteristic lifetime of an intermittent accretion disk that launches a pair of jets in the JJEM is $\simeq 0.01 -0.3 \s$ (e.g., \citealt{Soker2024Keyhole}), a time scale that is about equal or shorter than the viscous timescale of the accretion disk, which is its relaxation timescale (e.g., \citealt{Soker2024Keyhole}). The implication is that the accretion disk has no time to fully relax. Considering that the gas that feeds the accretion disk has large fluctuations in its properties, the two sides of the accretion disk are born unequal and have no time to relax. Such a disk will likely launch two opposite jets unequal in power and opening angle \citep{Soker2024CounterJet}.  

Consider the very early time after the shock bounces and stalls at $r \simeq 100-150 \km \s^{-1}$. At the period of $t_{\rm b} \lesssim 0.2 \s$, where $t_{\rm b}$ is the time measured from shock bounce, the mass accretion rate onto the very young central object, a bloated NS star, is $\dot M_{\rm acc} \gtrsim 1 M_\odot \s^{-1}$ (e.g., \citealt{Mulleretal2017, Varmaetal2021, Burrowsetal2024}). Due to its still lower mass and large radius, the escape velocity from the NS is lower than at the later phases of the explosion process, and we scale the jets' velocity by $ v_{\rm j}=5\times 10^4 \km \s^{-1}$; this is an expected minimum velocity, as it might be somewhat larger, up to $\simeq 10^5 \km \s^{-1}$. 
Consider a short accretion period through an accretion disk that lives for $\Delta t_1 \simeq 0.05 \s$. As discussed above, the disk doesn't have time to relax, and one jet can be much more powerful than the counter jet. We assume the more powerful jet carries much more momentum than the counter jet. This one jet carries a fraction $f_{\rm j1} \simeq 0.1$ of the accreted mass, while the counter jet carries a fraction $f_{\rm j2} \ll f_{\rm j1}$. The momentum and energy of the pair of jets are 
\begin{equation}
\begin{split}
p_{\rm j12} & =  250
\left( \frac{f_{\rm j1}-f_{\rm j2}}{0.1} \right)  
\left( \frac{\Delta t_1}{0.05\s} \right)  
\left( \frac{\dot M_{\rm acc}}{1 M_\odot \s^{-1}} \right) \\ & \times 
\left( \frac{v_{\rm j}}{5 \times 10^4 \km \s^{-1}} \right) 
M_\odot \km \s^{-1}, 
\label{eq:p1jet}
\end{split}
\end{equation}
and 
\begin{equation}
\begin{split}
E_{\rm j12} & =  1.24 \times 10^{50}
\left( \frac{f_{\rm j1}+f_{\rm j2}}{0.1} \right)  
\left( \frac{\Delta t_1}{0.05\s} \right)  \\ & \times
\left( \frac{\dot M_{\rm acc}}{1 M_\odot \s^{-1}} \right) 
\left( \frac{v_{\rm j}}{5 \times 10^4 \km \s^{-1}} \right)^2 
\erg, 
\label{eq:E1jet}
\end{split}
\end{equation}
respectively. 

The observed NS kick velocity distribution ranges from very slow to $v_{\rm NS} \simeq 1000 \km \s^{-1}$, with two peaks at $\simeq 80 \km \s^{-1}$ and $\simeq 500 \km \s^{-1}$ (e.g., \citealt{Igoshev2020}). For a typical NS mass of $M_{\rm NS}=1.4 M_\odot$ the momenta of most NSs are in the range $p_{\rm NS} \simeq 100-700 M_\odot \km \s^{-1}$. 
The highly asymmetrical and powerful jet pair at a very early time can account for this natal kick; we term it the kick by early asymmetrical pairs (Kick-BEAP) mechanism. The tug-boat mechanism can act later and change the value of the kick velocity that the kick-BEAP mechanism imparted. 

 \citet{Kondratyevetal2024} consider a kick imparted by launching one pair of jets with unequal jets in the magnetorotational supernova explosion model. The magnetorotational supernova explosion model requires a rapid pre-explosion corotation and, therefore, applies only to rare cases.  

Our suggested kick-BEAP mechanism imparts a large natal kick velocity to the NS at the beginning of the explosion process before the NS launches most jittering jets and, therefore, has the following properties and consequences. 
\begin{enumerate}
    \item There might be two or even three pairs of jets that can impart large kick velocity at $t_{\rm b} \la 0.2 \s$. 
    \item Because the core is still intact, the strong early jets can lead to the nucleosynthesis of elements from silicon to iron group with high asymmetrical distribution in the opposite direction to the natal kick velocity. This is the focus of a future study. 
    \item For the same reason, these early jets will all be choked inside the core and will not shape ears and bubbles in the CCSNR. Their marks are the kick velocity and the asymmetrical ejecta of some elements. 
    \item The NS launches the later jets, including the last pairs of jets with large marks on the CCSNR morphology, while it is already moving relative to the center of the collapsing core. This can explain the avoidance of small angles between the main jet axis and the kick velocity. We elaborate on this below. 
        \item   Because the kick-BEAP is along the first jet-pair, which is along the direction of the angular momentum of the accreted gas, at the end of the kick-BEAP operation, the spin and kick direction are aligned. In cases where the angular momentum of the following accretion episodes adds to a low value, the spin-kick alignment remains. This explains some CCSNe with spin-kick alignment, e.g., the pulsars in the CCSNRs S147 (e.g., \citealt{Yaoetal2021}) and Vela (e.g., \citealt{Noutsosetal2012}). However, later accretion episodes might, in some CCSNe, change the spin direction, leading to a misalignment (see some cases listed by, e.g., \citealt{Noutsosetal2012}). 
     We note that the neutrino-driven mechanism has difficulties explaining spin-kick alignment (for the most recent study, see \cite{SykesMuller2024}). 
 \end{enumerate}

Consider then that after the kick-BEAP phase, at $t_{\rm b} \gtrsim 0.2 \s$, the NS moves at a velocity of $v_{\rm NS} \approx 50-500 \km \s^{-1}$ relative to the center of mass of the pre-collapse core. We are interested in the final energetic pair of jets, the pair that is likely to shape the main jet axis \citep{Soker2024Keyhole}. The mass that the NS accretes at the final accretion phases of the explosion process, about a second to a few after the shock bounce, originates from a radius of $r_{\rm f} \simeq 3000 \km$ (e.g., \citealt{ShishkinSoker2021}). The impact parameter, namely the distance of the accreted mass from the trajectory of the NS, is in the range $0 \le b \le r_{\rm f}$. The specific angular moment that the kick velocity adds to an accreted parcel of gas with an impact parameter $b$ is perpendicular to the kick velocity with a magnitude of   
\begin{equation}
j_{\rm k}  = 3.75 \times 10^{15}
\left( \frac{b}{1500 \km} \right)  
\left( \frac{v_{\rm NS}}{250 \km \s^{-1}} \right)  \cm^2 \s^{-1}. 
\label{eq:jkick}
\end{equation}
This value is of the order of magnitude of the specific angular momentum fluctuations in the core convective zones (e.g., \citealt{ShishkinSoker2021,ShishkinSoker2022}).
Therefore, $\vec{j}_{\rm k}$ that is perpendicular to $\vec{v}_{\rm NS}$ prevents small angles between the kick velocity and the jet-main axis, and explains this finding (e.g., \citealt{BearSoker2023RNAAS}).

\section{Summary}
\label{sec:Summary}

We revealed a point-symmetric structure of two pairs of ears and one pair of clumps in the CCSNR Puppis A (Figure \ref{fig:PuppisAFig1}). The centers of the three symmetry lines that connect the two ears or clumps of each pair are within the general region between the present location of the NS remnant and its calculated location at the explosion time. 
Point-symmetric morphologies of CCSNRs, as we found here for Puppis A, are compatible with the expectation of the JJEM. Therefore, our findings further support the JJEM. 
Other shaping processes, like instabilities, interaction with a CSM, and the ISM, cannot account for the observed point-symmetric morphologies of CCSNRs (e.g., \citealt{SokerShishkin2024Vela}). In Section \ref{sec:PlanetaryNebulae}, we compared the point-symmetric morphology of Puppis A with three multipolar planetary nebulae that researchers consider to have been shaped by two or more pairs of jets. The similarities between Puppis A's point-symmetric morphology and the three planetary nebulae (Figure \ref{fig:PuppisA4PNe}) further solidify our claim that energetic jets shaped the point-symmetric morphology of Puppis A. According to the JJEM, these jets are part of $\approx 5$-$30$ pairs of jets that explode CCSNe (e.g., \citealt{Soker2025Learning}).   

The elongated morphological features of the `corridor' and the line connecting the ejecta-rich clump with the `tongue' (dashed lines on Figure \ref{Fig:4panels}) require deeper observations and analysis to explore their properties and whether jets also shaped their structures, as we suggest in this study.     

In this study, we also focused on the solid dipole structure of Puppis A. Its prominent components are the X-ray and radio bright `saddle' with a sharp edge on the northeast and the much fainter and diffuse other side. A possible ear, the `P-ear(?)', and the tongue (Figure \ref{Fig:4panels}) are the prominent components of the other dipole side of the saddle. We point out a similar, but not identical, dipole structure in the SNR N49 that we will study in a future paper (for images of N49, see, e.g., \citealt{Bilikovaetal2007, Ghavametal2024, Zhouetal2019}). The corridor in Puppis A that extends from one side to the other implies that an internal process shaped the dipole structures. Even if it is an absorbing material that causes the flatter radio spectrum, its structure correlation with the dipole axis requires an explanation. A CSM or an ISM cloud on the dense side of an SNR can compress that side but cannot form filaments and structures extending to the other side. Also, the NS kick velocity in Puppis A is opposite to the side of the dense part, the saddle (Figure \ref{Fig:4panels}). 

Based on the dipole structure of Puppis A and its relation to the kick velocity direction, we proposed (Section \ref{sec:kick}) a mechanism in the frame of the JJEM to impart a natal kick to the NS. In this kick-BEAP mechanism, the very young NS, age of $\lesssim 0.2 \s$, launches a pair of jets where one jet is much more energetic than the counter jet. This takes place when the accretion rate is very high. Momentum conservation implies that the NS recoils in the opposite direction of the much more energetic jets. Such highly unequal jets in a pair in many, but not all, cases is one of the expectations of the JJEM \citep{Soker2024CounterJet}. The more energetic jet in the pair compresses a dense side to the SNR while the NS acquires a natal kick velocity in the opposite direction. The kick-BEAP mechanism allows for the tug-boat mechanism, which occurs at a little later time in the explosion, to operate as well, but it is not necessary as the kick-BEAP might account for a large range of kick velocities (Eq. \ref{eq:p1jet}). 

Our study further supports the JJEM as the primary, or even sole, explosion mechanism of CCSNe and adds to the role of jets by introducing the novel kick-BEAP mechanism.

\begin{acknowledgement}

We thank Martin Mayer for advice regarding eROSITA data and useful comments, and an anonymous referee for helpful comments and the support. 

We acknowledge the use of NASA's \textit{SkyView} facility \\ (https://skyview.gsfc.nasa.gov) located at NASA Goddard Space Flight Center.

This work made use of Astropy:\footnote{http://www.astropy.org} a community-developed core Python package and an ecosystem of tools and resources for astronomy \citep{astropy:2013, astropy:2018, astropy:2022}.
We also made use of Matplotlib \citep{Hunter:2007} and NumPy \citep{harris2020array} in our image processing.

This work makes use of data from eROSITA, the soft X-ray instrument aboard SRG, a joint Russian-German science mission supported by the Russian Space Agency (Roskosmos), in the interests of the Russian Academy of Sciences represented by its Space Research Institute (IKI), and the Deutsches Zentrum für Luft- und Raumfahrt (DLR). The SRG spacecraft was built by Lavochkin Association (NPOL) and its subcontractors, and is operated by NPOL with support from the Max Planck Institute for Extraterrestrial Physics (MPE). The development and construction of the eROSITA X-ray instrument was led by MPE, with contributions from the Dr. Karl Remeis Observatory Bamberg \& ECAP (FAU Erlangen-Nuernberg), the University of Hamburg Observatory, the Leibniz Institute for Astrophysics Potsdam (AIP), and the Institute for Astronomy and Astrophysics of the University of Tübingen, with the support of DLR and the Max Planck Society. The Argelander Institute for Astronomy of the University of Bonn and the Ludwig Maximilians Universität Munich also participated in the science preparation for eROSITA.

The eROSITA data shown here were processed using the eSASS software system developed by the German eROSITA consortium.

This publication makes use of data products from the Wide-field Infrared Survey Explorer, which is a joint project of the University of California, Los Angeles, and the Jet Propulsion Laboratory/California Institute of Technology, funded by the National Aeronautics and Space Administration.

We used the PNIC: Planetary Nebula Image Catalogue composed by Bruce Balick \\(http://faculty.washington.edu/balick/PNIC/)

\end{acknowledgement}

\paragraph{Funding Statement}
A grant from the Pazy Foundation supported this research.

\end{document}